# Changes in Retirement Savings During the COVID Pandemic

| Elena Derby | Lucas Goodman | Kathleen Mackie | Jacob Mortenson[†] |
|---|---|---|---|
| Joint Committee on Taxation, U.S. Congress | Office of Tax Analysis, U.S. Treasury Department | Joint Committee on Taxation, U.S. Congress | Joint Committee on Taxation, U.S. Congress |

**April 2022**

### Abstract

This paper documents changes in retirement saving patterns at the onset of the COVID-19 pandemic. We construct a large panel of U.S. tax data, including tens of millions of person-year observations, and measure retirement savings contributions and withdrawals. We use these data to document several important changes in retirement savings patterns during the pandemic relative to the years preceding the pandemic or the Great Recession. First, unlike during the Great Recession, contributions to retirement savings vehicles did not meaningfully decline. Second, driven by the suspension of required minimum distribution rules, IRA withdrawals substantially declined in 2020 for those older than age 72. Third, likely due to the partial suspension of the early withdrawal penalty, employer-plan withdrawals increased for those under age 60.

*JEL Codes: G51, H24, J32*

[†]Corresponding author: Jake.Mortenson@jct.gov;
502 Ford House Office Building, Washington, DC 20515; (202) 225-3261
This research embodies work undertaken for the staff of the Joint Committee on Taxation, but as members of both parties and both houses of Congress comprise the Joint Committee on Taxation, this work should not be construed to represent the position of any member of the Committee. The views expressed in this paper are not necessarily those of the U.S. Department of the Treasury. The authors declare that they have no relevant financial interests that relate to the research described in this paper. The authors thank Thomas Barthold, Sally Kwak, and Michael Love for providing comments.

*INTRODUCTION*

The COVID-19 pandemic brought about a fast and severe global economic decline. By the end of March 2020, the United States economy had lost over 13 million jobs, compared to under 9 million jobs lost during the Great Recession (Cajner, et al, 2020). Although some of the effects were mitigated by government interventions – increased unemployment insurance (UI), stimulus checks, government assistance to private entities, and other public assistance – changes in asset prices, employment, and consumption were substantial.

In this paper, we use tax data to document how retirement savings – specifically, contributions and withdrawals – changed during the COVID-19 pandemic in comparison to the Great Recession (from 2008 to 2010). Retirement savings behavior might plausibly have been affected both by the pandemic (and related recession) itself, as well as by the policy actions that Congress took in response to the pandemic.

We find little change in contributions in 2020, which increased from the previous year at a rate in line with recent trends, a stark difference from the drop in contributions that occurred during the Great Recession (Argento, Sablehaus, and Bryant, 2015; Goodman, Mackie, Mortenson, and Schramm, 2021). This may have occurred because the effects of the COVID-19 pandemic were disproportionately worse for workers at the bottom of the income distribution, who save at much lower rates than middle- and high-income earners, while the earnings shocks during the Great Recession affected middle- and high-income workers to a greater degree than during COVID-19 (Larrimore, Mortenson, and Splinter, 2022).

Conversely, withdrawal patterns in 2020 were meaningfully different from prior years, falling for older individuals and increasing for younger people. Both patterns were influenced by policy changes that Congress made in response to the pandemic. First, the requirement for older individuals to take certain withdrawals (known as required minimum distributions, or RMDs) was suspended in 2020. We find that people responded to this policy by sharply reducing their withdrawals, much as they did in response to the suspension of those same requirements in 2009. Second, Congress granted broad (but not complete) relief from the 10 percent penalty that applies to most withdrawals made by individuals under age 59½. Using bunching methods, we find clear evidence that some individuals responded to this policy by taking large (approximately $100,000) withdrawals. At the same time, we also find evidence that many people (at least, those near age 59½) took withdrawals consistent with the penalty remaining in place. It is also likely that some of the increase in withdrawals by working-age individuals can be attributed to the economic environment – especially increased job separation – though we are unable to fully test this hypothesis with the data currently available.

Our research builds on prior research examining the effects of the COVID-19 pandemic – and subsequent federal policy changes – on economic activity. Prior research has examined changes in unemployment and inequality (Bartik, et al., 2020; Cajner, et al., 2020; Clark, Lusardi and Mitchell, 2020; Coibion, Gorodnichenko and Weber, 2020; Guerrieri, et al., 2020), consumer spending (Chetty, et al., 2021), and early retirement (Goda, et al., 2021; Bui, et al., 2020; Davis, 2021) during the pandemic. Others have studied the effects of both federal and state government interventions during the pandemic including stay-at-home orders (Forsythe, et al., 2020), and federal assistance programs like unemployment insurance and stimulus payments (Baker, et al., 2020; Chetty, et al., 2021, Larrimore et al., 2022).

This paper also contributes to the body of work exploring the effect of economic shocks and policy changes on retirement savings. On the topic of RMDs, Brown, Poterba and Richardson (2017) find that the suspension of RMDs in 2009 resulted in a large decrease in withdrawals for TIAA-CREF retirement savings participants, particularly among relatively wealthier individuals with large balances, and those with longer retirement horizons (people closer to the starting age for RMDs). Mortenson, Schramm and Whitten (2019) find that 32 to 52 percent of individuals subject to RMDs would prefer to take a withdrawal below the required minimum, but that even when the RMD is suspended, some individuals take withdrawals at the "phantom" (i.e., not in effect for that year) RMD threshold. Berdshadker and Smith (2005) also estimate – using Current Population Survey (CPS) data – that about 45 percent of individuals with positive account balances do not take withdrawals until they are required to do so by law.

Work exploring the effect of the early withdrawal penalty is sparser than the RMD literature. Goda, et al. (2011) have also measured individuals' responsiveness to the penalty by estimating the increase in withdrawals at the threshold age of 59½, finding that the probability of a withdrawal increases by 93 percent at this age. Stewart and Bryant (2021) similarly find that the early withdrawal penalty (and RMDs) substantially influence withdrawal timing. Several studies have found that the likelihood of taking early withdrawals increases significantly with negative shocks including divorce and job loss or wage reduction (Argento, Sablehaus, and Bryant, 2015; Amromin and Smith, 2003; Goodman, et al., 2021; Brady, 2019), the latter of which applies directly to the COVID-19 employment shocks. In relation to the Great Recession, Argento, Sablehaus, and Bryant (2015) also find that contributions to defined contribution (DC) plans declined significantly, and Goodman, et al. (2021) find that net contributions did not recover their inflation-adjusted 2007 levels until 2014. Both papers also find that early withdrawals increased during the Great Recession, although to a smaller extent than the change in contributions. Goodman, et al. (2021) find that the share of contributions made by working age Americans that exited the retirement saving system increased modestly during the Great Recession, from a base rate of around 22 percent to 26 percent in 2009.

There has also been a focus on the effects of economic downturns on early retirement. Coile and Levine (2017) find that economic downturns cause an increase in the percent of people who decide to retire, equivalent to the effect of a negative health shock or retirement incentives. Neumark and Button (2014) also find that age discrimination was a problem for workers nearing retirement during the Great Recession. However, Goda, et al. (2016) find that during the Great Recession the probability of retiring early actually went down, which may have been a reaction to stock market conditions at that time. More recently, Goda, et al. (2021) find that despite evidence suggesting that the COVID-19 pandemic caused an increase in early retirement, there was no significant increase in applications for Social Security benefits, which is in line with our findings, as we do not see an increase in retirement withdrawals for individuals between the ages of 60 and 70.

## I.   *Background and Policy Changes*

Under the Coronavirus Aid, Relief, and Economic Security (CARES) Act of 2020 the rules for distributions and withdrawals from retirement accounts changed in two ways: required minimum distributions were temporarily suspended and "coronavirus-related" early withdrawals were exempted from early distribution penalties. There were also changes to rules regarding rollovers, loan limits and repayments, and partial plan terminations, but we omit these from our discussion as they are beyond the scope of our analysis.[1]

In general, owners of IRAs and DC accounts must begin taking withdrawals — known as Required Minimum Distributions (RMDs) — from their accounts in the year in which they reach age 72. These withdrawals are equal to a specified fraction of the prior-year balance. In the CARES Act, Congress suspended RMDs for calendar year 2020, meaning that no such withdrawals were required during that year.[2] Congress granted similar relief during the Great Recession, suspending RMDs in 2009 (JCT, 2020; Topoleski, 2020).

Non-rollover withdrawals from IRAs and DC accounts made prior to age 59½ are generally subject to a 10 percent penalty, in addition to (for traditional, non-Roth accounts) being subject to ordinary income tax at the time of withdrawal. In the CARES Act, Congress created an exception from the early withdrawal penalty for "coronavirus-related" withdrawals from IRAs or eligible retirement plans (such as employer provided 401(k) and 403(b) plans) in 2020. This exception was quite broad; it applied to withdrawals up to $100,000 received by any individual who experienced self-attested economic or health hardships during 2020. Second, the law permitted the recognition of coronavirus-related withdrawals over three years, rather than at the time of this withdrawal, which provided taxpayers with potentially needed liquidity, allowed taxpayers to take advantage of the time value of money, and provided the opportunity to smooth their income and minimize tax payments under a progressive tax schedule. Third, the law permitted individuals to "recontribute" coronavirus-related withdrawals at any point in 2020, 2021, or 2022 by making contributions back into the distributing account. Any such repayments were to be treated as if they were never withdrawn (i.e., there would be no income inclusion).

## II. Data

The data used to generate the estimates, figures, and tables in this paper are drawn from administrative tax records in the United States. These include data compiled from Form 1040, which is filed by the taxpayer, and many information returns, which are filed by third parties, including Forms W-2, 1099-R, 1099-SSA, and 5498. The base data are in the form of an unbalanced individual-level panel including five percent of all individuals in the U.S. tax population.[3] While the panel is unbalanced, it is representative of the population of taxpayers within each year from 2003 to 2020, and individuals can enter the panel by immigration, birth, or receiving a tax form, and they can exit the panel by emigration, death, or failing to receive a tax form. These base data have been used in several prior papers (Mortenson, Schramm, and Whitten, 2019; Goodman, et al., 2021; Larrimore, Mortenson, and Splinter, 2022).

---

[1] Specifically, the CARES Act allowed plans to increase the section 72(p) loan limit from $50,000 to $100,000 for certain loans and allowed plans to elect to suspend certain loan payments. The CARES Act also protected employers from inadvertently generating a "plan termination" if a large share of its workforce separated from employment.
[2] The SECURE Act changed the threshold age from 70.5 to 72. This change would have taken effect in 2020 if Congress had not passed the RMD holiday in the CARES Act.
[3] The tax population – those individuals who appear on a federal income tax return (1040) or information return (e.g., Form W-2) includes roughly 98 percent of the Census resident population in each year (Larrimore, Mortenson, and Splinter, 2020).

This paper relies most heavily on the following pieces of information retrieved from administrative data: date of birth and death from the Social Security Administration's (SSA) DM-1 file, IRA contributions from Form 5498, wages and deferred contributions from Form W-2 (wages), Social Security benefits from Form 1099-SSA, and retirement distributions from Form 1099-R. All dollar amounts are adjusted to 2020 price levels using chained CPI.

Form 1099-R allows us to distinguish between distributions made from IRAs, Roth IRAs, and workplace pension plans. However, within the category of workplace pension plans, we are not able to distinguish between distributions made from defined contribution (DC) plans and those made from defined benefit (DB) plans. In a prior paper Goodman, et al. (2021) applied an algorithm to classify 1099-R distributions as DB or DC based on various factors. However, this algorithm is unable to classify data from 2020 because it requires several years of "burn-out" (i.e., the last several years of data are used to classify earlier distributions but cannot be classified themselves). Therefore, when analyzing distributions from workplace pension plans, we combine both types of distributions.

## III. *Retirement Savings Changes*

CONTRIBUTIONS

We begin by examining the mean dollar amount of contributions made by individuals to employer-sponsored DC accounts and IRAs. In **Figure 1** we plot average contributions by age, separately by year. The upper panel (**Figure 1a**) displays mean contributions across the age distribution during the Great Recession, from 2008 to 2010, and the lower panel for 2018 to 2020.

We find significantly different patterns at the onset of the pandemic versus the Great Recession. Contributions increased for those under 65 in 2020, mostly on-trend relative to the two (non-pandemic) years immediately preceding. This contrasts to the experience of the Great Recession, where there is a clear reduction in retirement contributions made by those under 60 in 2009 and 2010 relative to 2008, and an increase in contributions made by those over 65 in 2010 relative to 2008 and 2009.[4]

There are several explanations for the different experience of the COVID-19 recession. First, retirement savings are concentrated in the upper half of the wage income distribution: in 2019, approximately 90 percent of contributions to employer-sponsored defined contribution plans were made by workers in the top two quintiles of the wage distribution. As shown in Larrimore, Mortenson, and Splinter (2022), the labor income shocks of the COVID-19 recession were concentrated in the lower half of the wage earnings distribution relative to the Great Recession, reducing the scope for negative income shocks to affect retirement contributions in the aggregate.

---

[4] Contribution patterns by age in 2007 did not meaningfully differ from those in 2008.

Second, the pandemic and associated closures reduced the marginal utility of consumption, especially in the upper half of the income distribution. As Chetty, et. al. (2020) find, consumer spending among individuals in the bottom quartile of the income distribution in the United States remained about the same after June 2020; however, consumption among those in the top quartile of income dropped by about 13 percent. This reduction in the opportunity cost of contributing to a retirement account makes retirement saving relatively more attractive, pushing against any reductions in saving caused by negative labor income shocks.

To explore this further, we show the change in contributions (relative to the prior year) as a function of prior year labor earnings, defined as wage earnings plus unemployment insurance.[5] **Figure 2** displays the mean change in contributions to employer-sponsored DC plans and IRAs in 2008, 2009, 2019, and 2020. A value of $0 in 2009 indicates that average contributions did not change from 2008 to 2009, whereas a value of $200 indicates contributions increased by $200 on average. Each series is displayed separately by prior-year earnings centile. To smooth each series, we aggregate individuals into bins five percentage points wide (ventiles). We break the top ventile into two pieces: one containing the 95$^{th}$-99$^{th}$ centiles, and another containing only the top centile. Individuals without prior year earnings are excluded from the sample, as are individuals younger than age 24 or older than age 69.

In each year, year-over-year changes in dollar amounts for the bottom quintile are small, consistent with this group not making large retirement contributions on average. However, the patterns farther up the distribution differ substantially for the years we examine. In 2008, during the beginning of the Great Recession, contributions tapered, especially at the very top: those in the top earnings centile decreased contributions (relative to 2007) by approximately $600 on average. The following year, 2009, saw an even sharper drop-off, except for the very top centile. This group increased contributions relative to 2008 by over $600 on average, roughly returning to 2007 levels. In 2019, the year prior to the onset of the COVID-19 pandemic, contributions increased relative to 2018 in a manner consistent with contributions increasing in earnings. The pattern in 2020 was much more similar to 2019 than 2009: the increase in contributions relative to the previous year remained flat for the bottom half of the distribution, increasing in earnings, with an even larger increase for the top decile. Top centile increases in 2009 and 2020 are likely partly a reflection of the (real) increases in the maximum contribution limit in those years, as many high-earnings individuals are bound by these limits.

Third, we observe different patterns is that the data years available are only comparable to varying degrees. For example, stock market conditions differed dramatically between the Great Recession and the COVID-19 pandemic. In October of 2007, the S&P 500 attained a high of 1,565, after which it tumbled over the course of 17 months to achieve a nadir of 677 in March of 2009, before embarking on a steady recovery. It did not re-attain its October 2007 value until April 2013. By contrast, the COVID-19 pandemic saw only a very brief downturn: On January 31, 2020, the S&P 500 closed at 3,226. It then fell to a low of 2,237 on March 23, 2020, after which it recovered rapidly. It re-attained its January 31, 2020, value by June 8, 2020. Given the well-known recency bias in investment decision-making (Tversky and Kahneman, 1973) – especially by retail investors who are responsible for the bulk of retirement savings (Nofsinger and Varma, 2013) – the longer bear market during the Great Recession may have had a larger effect on contributions than the V-shaped market during the COVID pandemic.

---

[5] Wage earnings are defined as box 5 from Form W-2, Medicare Wages.

WITHDRAWALS:

**Figure 3** plots mean withdrawals (non-rollover or conversion distributions) from retirement accounts by age, from 2008 to 2010 (**Figures 3a** and **3b**) and from 2018 to 2020 (**Figures 3c** and **3d**). **Figure 3a** shows the change in mean withdrawals from IRAs by age during the Great Recession, from 2008 to 2010. **Figure 3b** does so for non-IRA accounts -- primarily employer sponsored defined contribution (DC) and defined benefit (DB) plans – from 2008 to 2010. **Figures 3c** and **3d** are analogous figures for the COVID recession, comparing 2018-2019 with 2020.

Several facts emerge. First, withdrawals from IRAs fell substantially in 2009 and 2020 for ages that would otherwise be subject to the required minimum distribution (RMD). Second, we see no large changes in withdrawals in those same years for individuals between ages 60 and 70, people who were relatively unaffected by the major policy changes. Third, while withdrawals from workplace DB and DC pension plans remained similar across years during the Great Recession, there is a noticeable increase in early withdrawals for pre-retirement-age cohorts in 2020 relative to the previous two years.

The effects of the 2020 RMD suspension are consistent with the same 2009 policy change. In **Figure 4**, we graph the mass of withdrawals from IRAs taken by 75-year-olds, scaled by their prior year balance, with reference to their RMD (or what the RMD would have been in the absence of the rule suspension).[6] The top panel (**Figure 4a**) plots withdrawal levels in 2009 (line) along with the same withdrawal levels in 2008 (dots), and the bottom panel (**Figure 4b**) plots withdrawal levels in 2020 (line) and 2019 (dots) for comparison. In both cases, there is a substantial shifting of mass from exactly the RMD to zero. Furthermore, there remains some mass at the RMD in 2009 and 2020, potentially reflecting a mix of inertia, a perception of the RMD as "guidance", or withdrawals taken during the year prior to the policy change (Brown, Poterba, and Richardson, 2017; Mortenson, Schramm, and Whitten, 2019).

The population near retirement age is also of interest, as there is a general concern that the pandemic and related recession induced people to retire earlier than they otherwise would – a concern that is corroborated by empirical evidence on labor force participation (Davis, 2021; Schwartz and Marcos, 2021; Van Dam, 2021; Hsu, 2021). In our data, we do not see any obvious increase in retirement withdrawals for those aged 60-70. Furthermore, Goda, et al. (2021) find that the number of applications to file for Social Security retirement benefits also remained largely unchanged during the pandemic. Thus, based on our findings and those in Goda, et al. (2021), we see no evidence of near-retirement age individuals decumulating their retirement wealth — whether in the form of Social Security, pensions, or IRAs — earlier than they otherwise would. If the pandemic caused early retirement, then these early retirees must be financing their consumption through some other source. Perhaps, as suggested by Goda, et al. (2021), increased unemployment insurance and other forms of pandemic assistance are allowing these individuals to finance their early retirement without needing to dip into their retirement wealth.

---

[6] We display the results for 75-year-olds here, but the results are similar for other age groups subject to RMDs.

The third finding — the increase in workplace pension withdrawals for those of working age — likely reflects both a response to policy changes and direct effects of the pandemic. We begin by showing that the policy changes were meaningful. In **Figure 5**, we plot a time series of the total withdrawals from pensions and IRAs (excluding rollovers, Roth conversions, etc.) and total penalized withdrawals for those under age 59½.[7] While total withdrawals increased by $60 billion (or 25 percent) in 2020, the total amount of penalized distributions fell by nearly 50 percent. Thus, it appears that take-up of the penalty suspension was quite high.

We find much smaller take-up of the CARES Act provision that allows for the income recognition of withdrawals to be spread over three years. In particular, we find that only 1.5 percent of tax units that took withdrawals from retirement accounts reported taxable pension and IRA amounts approximately equal to 1/3 of total withdrawals from Form 1099-R. Perhaps, take-up was low for this provision because the hassle costs of needing to track this information to subsequent tax returns exceeded the value of the deferral of liability or other tax savings.[8]

In **Figure 6**, we show clear evidence that some pre-retirement age individuals were responding to a combination of these policies. Recall that the penalty suspension and the three-year recognition rule can apply to up to $100,000 of withdrawals for any given individual. **Figure 6** shows that there is indeed bunching at this threshold in 2020: approximately 165,000 individuals (between the ages of 20 and 58) took a withdrawal within $500 of $100,000, far larger than any nearby bins of the same width. This is not driven by round-number bunching, as we see no analogous pattern in 2019. Under the extreme assumption that these individuals would have not taken any withdrawals at all in 2020 in the absence of the policy changes, then these bunchers alone would account for approximately $17 billion of the $65 billion increase in withdrawals from 2019 to 2020 for this age group. While we cannot test this assumption directly, we believe that it is a plausible approximation: only 10 percent of these individuals took a withdrawal of any amount in 2019. Taken together, this suggests that a non-trivial portion of the increase in withdrawals from 2019 to 2020 was driven by policy.

These bunchers could plausibly have been responding to any of the policy changes that applied to coronavirus-related withdrawals. We interpret this bunching to mostly be a response to the penalty suspension for two reasons. First, we find that the probability of paying a penalty was substantially lower for those in the bin nearest to $100,000 in withdrawals relative to those in nearby bins. Second, only 7 percent of these bunchers appeared to take-up the three-year recognition option, using the proxy described above (taxable withdrawals approximately equal to one third of total withdrawals).

While the evidence on bunching indicates that the relaxation of the early withdrawal penalty played a role in increasing withdrawals in 2020, we also find evidence that many individuals still felt constrained by the penalty, at least to some extent. We follow the methodology in Goda, et al. (2011) to examine the probability of taking withdrawals as a function of monthly age in the vicinity of 59½. Consider the blue (2019) series in each panel of Figure 7. Individuals with relative age 0 are those who reach age 59½ in January of 2020 – they have zero months in 2019 during which to take penalty-free withdrawals. As we move left in the graph, we move to even younger individuals (i.e., who reach age 59½ later in the year) who also had zero penalty-free months in 2019. But, as we move right, we move towards individuals who had more months of penalty-free withdrawals – up until we hit relative age 12 months, at which point the entire year is penalty-free. Indeed, we see changes in the slope of the probability of taking a non-IRA withdrawal (top panel) or an IRA withdrawal (bottom panel) at relative age 0 and 12 months, suggesting that the probability of taking a non-IRA withdrawal in any given year is increasing approximately linearly in the number of penalty-free months during that year.

---

[7] We approximate penalized distributions as retirement-related penalties reported on Form 1040 among people under age 59½ divided by 10 percent (the penalty amount).
[8] We currently lack the data to analyze the take-up of the CARES Act provision that allows for subsequent repayment of coronavirus-related withdrawals.

The key empirical test is whether these discontinuous changes in slope are present in 2020 as well. If people perceived the relief for coronavirus-related withdrawals as effectively eliminating the penalty for all withdrawals, then there should be no slope change; all individuals would have faced 12 months of penalty-free withdrawals.[9] Yet, we see very similar slope changes at 0 and 12 months in 2020 versus 2019. For the non-IRA series, the two years are indistinguishable, potentially reflecting the measurement error introduced by the fact that both series include DB withdrawals which are (in general) not subject to the penalty in either year. There is a more noticeable difference in the IRA series: the change in slope in 2020 is statistically significantly smaller than the change in slope in 2019. However, the difference is modest in magnitude: the change in slope shrinks by only 17 percent.

Taking **Figure 6** and **Figure 7** together suggests a nuanced, heterogeneous response to the relief granted to coronavirus-related withdrawals. **Figure 6** suggests that a select group of about 140,000 people responded to the policy in a sophisticated way, and the behavior of this group was large enough to affect aggregate withdrawals. Meanwhile, **Figure 7** suggests that most individuals took withdrawals in a manner consistent with the penalty remaining in place, either because they did not deem their distributions to be coronavirus-related (i.e., the penalty would have applied) or because they were unaware of the penalty suspension.

Finally, we note that it is likely that some of the increase in withdrawals to working-age persons in 2020 is not the result of policy, but rather the result of the economic environment. As shown in Goodman, Mackie, Mortenson, and Schramm (2021), job separation is associated with a substantial increase in the probability of working-age persons taking retirement withdrawals. This may be caused both by the income shock associated with job separation, but also the choice architecture that often makes withdrawals the default option, or easiest option, for those with accumulated DC balances in their previous job. Thus, the large increase in job separation that occurred in 2020 may have contributed to the increase in withdrawals in this period. In future research – once we are able to observe W-2 data from 2021, which are necessary for determining whether an individual had a job separation event – we intend to quantify the role of increased job separation on withdrawals in 2020.

---

[9] Given that the CARES Act passed at the end of March 2020, one might expect a kink at relative age 9 months (i.e., those attaining 59½ around March or April 2020) rather than relative age 12 months. But this should not affect the kink at relative age equal to zero months (i.e., those attaining age 59½ near the end of 2020).

## IV. Conclusion

The COVID-19 pandemic had many negative consequences that continue to persist through early 2022. The sharp increase in unemployment, supply-chain issues, increasing prices, and hesitancy of workers to return to the labor force all produced strains on the economy. Yet, due to the differential impacts of the pandemic on the workforce, aggregate contributions to retirement accounts remained largely the same, and potentially increased beyond what was expected for high-income workers. Withdrawals from employer-sponsored retirement accounts to working-age individuals did increase substantially during 2020, but we find evidence that this was driven at least to some extent by a response to relief granted to certain early withdrawals, including the elimination of the 10 percent early distribution penalty for coronavirus-related reasons. We expect that some portion of this increase was also driven by the increase in job separations created by the pandemic. Withdrawals from IRAs to those over 70 dropped precipitously, suggesting a strong response to the suspension of required minimum distribution rules. The source of the asymmetric response in IRAs and employer-sponsored plans to these two policy changes is unclear but is likely driven by the co-mingling of defined benefit withdrawals with defined contribution withdrawals in tax data reporting and the age-life cycle of IRA balances.

Bui, Truc Thi Mai, Patrick Button, and Elyce G. Picciotti (2020). "Early evidence on the impact of coronavirus disease 2019 (COVID-19) and the recession on older workers." *Public Policy & Aging Report* 30, no. 4: 154-159.

Cajner, Tomaz, Leland Dod Crane, Ryan Decker, Adrian Hamins-Puertolas, and Christopher Johann Kurz (2020). "Tracking labor market developments during the covid-19 pandemic: A preliminary assessment." Working paper.

Chetty, Raj, John N. Friedman, Nathaniel Hendren, Michael Stepner (2020). *How did COVID-19 and stabilization policies affect spending and employment? A new real-time economic tracker based on private sector data*. Cambridge, MA: National Bureau of Economic Research.

Clark, Robert L., Annamaria Lusardi, and Olivia S. Mitchell (2021). "Financial fragility during the COVID-19 Pandemic." In *AEA Papers and Proceedings*, vol. 111, pp. 292-96.

Coibion, Olivier, Yuriy Gorodnichenko, and Michael Weber (2020). *Labor markets during the COVID-19 crisis: A preliminary view*. No. w27017. National Bureau of Economic Research.

Coile, Courtney C., and Phillip B. Levine (2007). "Labor market shocks and retirement: Do government programs matter?" *Journal of Public Economics*, 91, no. 10: 1902-1919.

Davis, Owen (2021). "Employment and Retirement Among Older Workers During the COVID-19 Pandemic." Working paper. Schwartz Center for Economic Policy Analysis.

Forsythe, Eliza, Lisa B. Kahn, Fabian Lange, and David Wiczer (2020). "Labor demand in the time of COVID-19: Evidence from vacancy postings and UI claims." *Journal of Public Economics* 189: 104-238.

Goda, Gopi Shah, Emilie Jackson, Lauren Hersch Nicholas, and Sarah See Stith (2021). "The Impact of Covid-19 on Older Workers' Employment and Social Security Spillovers." No. w29083. National Bureau of Economic Research.

Goda, Gopi Shah, Damon Jones, Shanthi Ramnath, and U. S. Treasury (2016). "How Do Distributions from Retirement Accounts Respond to Early Withdrawal Penalties? Evidence from Administrative Tax Returns." *RRC Paper No. NB16-05. Cambridge, MA: National Bureau of Economic Research*.

Goda, Gopi Shah, John B. Shoven, and Sita Nataraj Slavov (2011). "What explains changes in retirement plans during the Great Recession?" *American Economic Review* 101, no. 3: 29-34.

Goodman, Lucas, Kathleen Mackie, Jacob Mortenson, and Heidi R. Schramm (2021). "The Evolution of Leakage and Retirement Asset Flows in the US." *National Tax Journal* 74, no. 3: 689-719.

Guerrieri, Veronica, Guido Lorenzoni, Ludwig Straub, and Iván Werning (2020). "Macroeconomic implications of COVID-19: Can negative supply shocks cause demand shortages?" No. w26918. National Bureau of Economic Research.10

**Figure 1: Mean DC and IRA Contributions by Age**

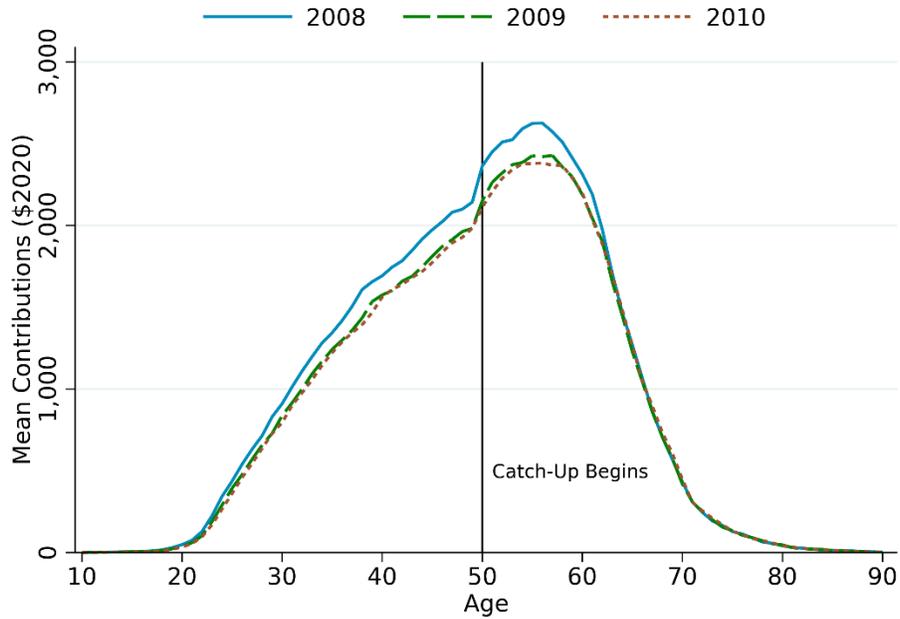

Figure 1a. 2008-2010

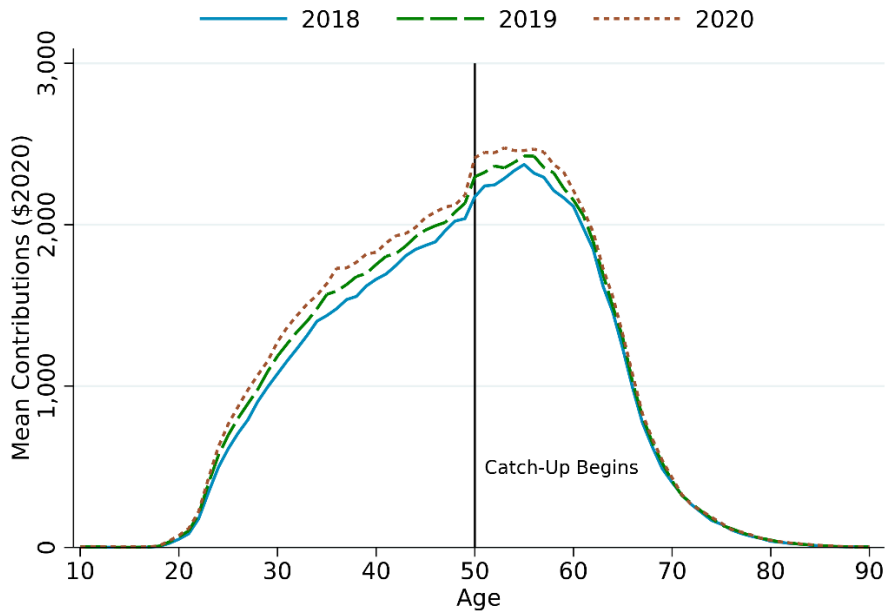

Figure 1b. 2018-2020

**Notes:** This figure plots mean contributions to workplace defined contribution plans and IRAs by age, separately by year. The vertical line highlights age 50, the age at which catch-up contributions begin. All dollar amounts are adjusted to 2020 price levels. The sample is representative of all individuals of a given age that appear on a tax return or an information return, including those who do not contribute. The data underlying the figure are a 5 percent random sample of IRS records derived from tax returns and information returns.

**Figure 2.: Mean Change in DC and IRA Contributions by Prior-Year Earnings Centile**

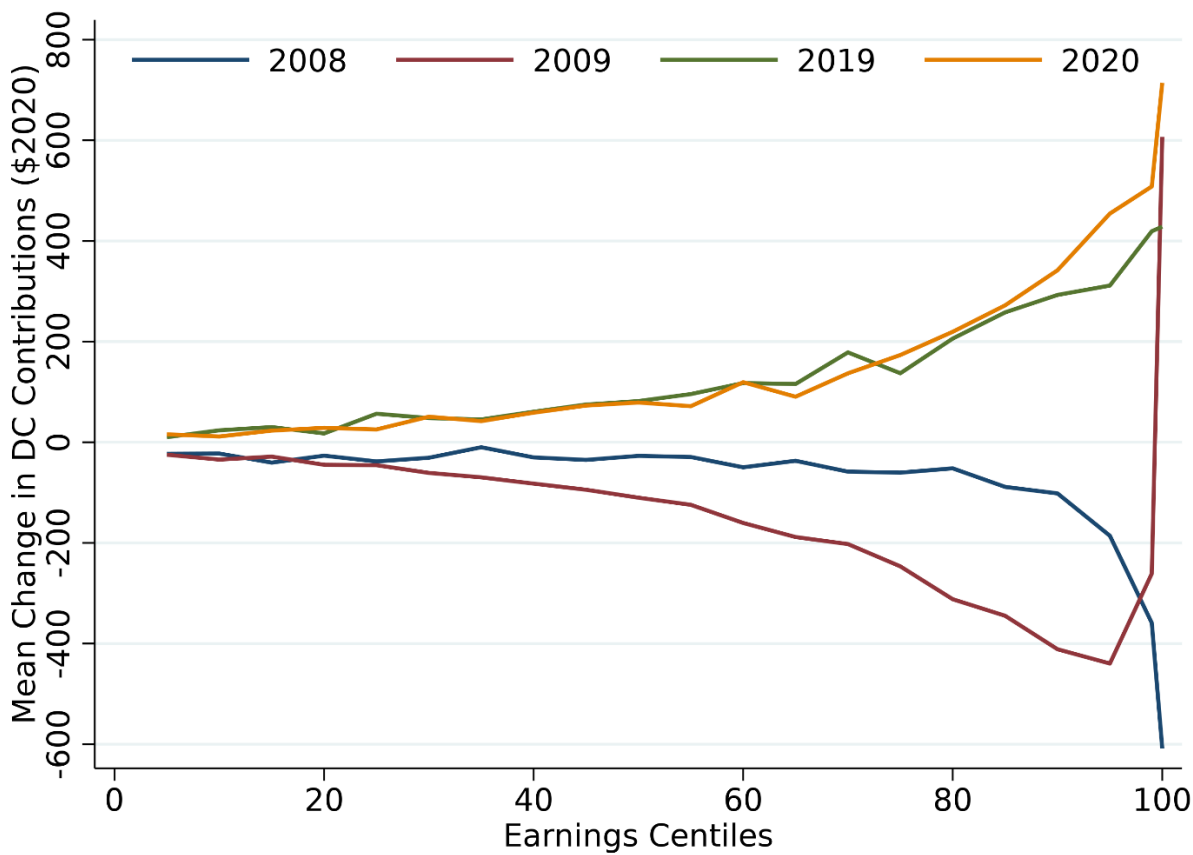

**Notes:** This figure plots the mean change in contributions to workplace defined contribution plans and IRAs, in 2020 dollars, from a "base year" to a "following year" as a function of earnings centiles in the base year. The legend indicates the series associated with each following year. Earnings are defined as W-2 wages plus unemployment compensation. The sample is representative of all individuals aged 24 to 69 with earnings in the prior year, including those who do not contribute. Centiles are combined into 5 centile-wide bins, except the top ventile, which is separately grouped as the 95th to 99th centiles and the top centile. The data underlying the figure are a 5 percent random sample of IRS records derived from tax returns and information returns.

**Figure 3: Mean Withdrawals from IRAs and Other Retirement Accounts, by Age**

**Figure 3a. IRA Withdrawals, 2008-2010**

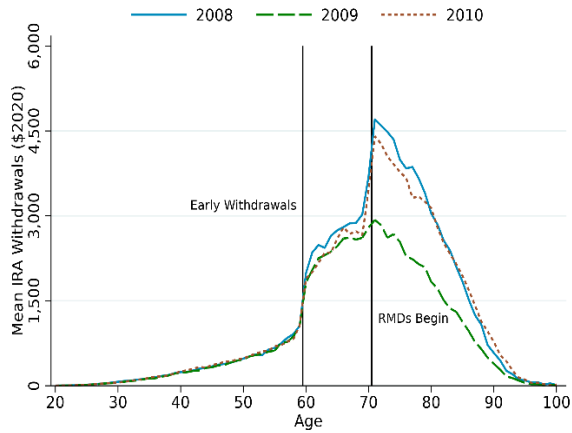

**Figure 3b. Non-IRA Withdrawals, 2008-2010**

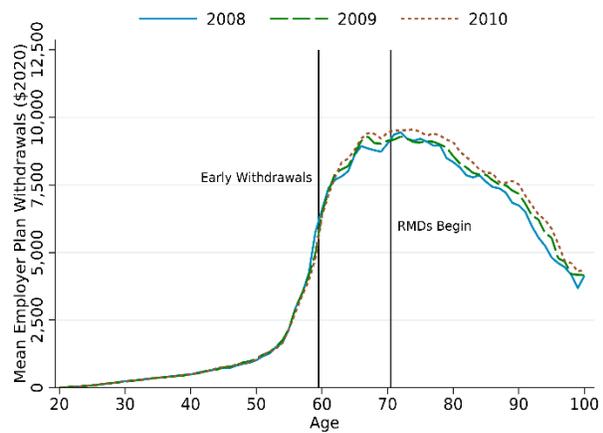

**Figure 3c.: IRA Withdrawals, 2018-2020**

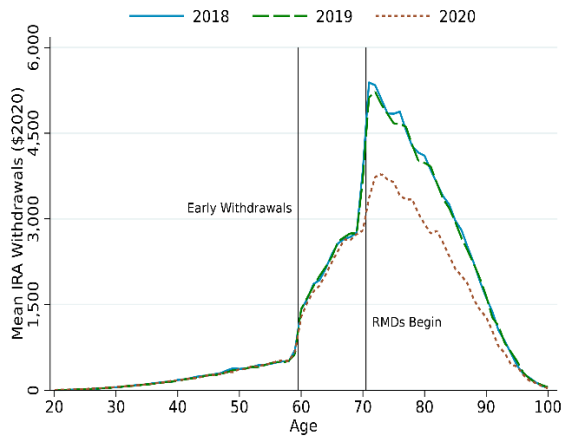

**Figure 3d: Non-IRA Withdrawals, 2018-2020**

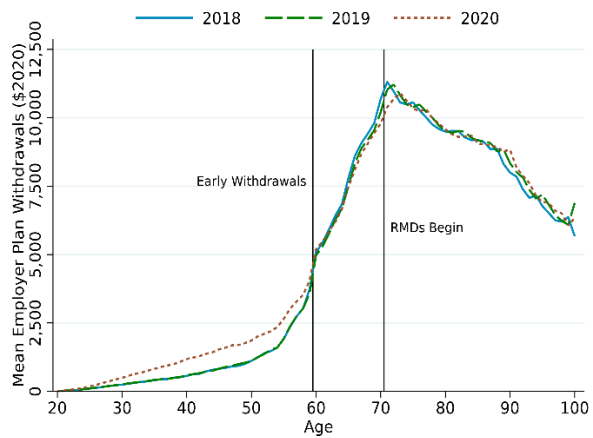

**Notes:** This figure plots mean withdrawals from IRAs (panels (a) and (b)) and workplace defined contribution and defined benefit plans (panels (b) and (d)) as a function of age, separately by year. Withdrawals are derived from Form 1099-R and exclude rollovers and Roth conversions. All dollar amounts are adjusted to 2020 price levels. The sample is representative of all individuals of a given age that appear on a tax return or an information return, including those who do not take a withdrawal. The data underlying the figure are a 5 percent random sample of IRS records derived from tax returns and information returns.

**Figure 4: Counts of IRA Withdrawals as a Percent of Prior Year Balance for 75-year-olds**

**Figure 4a. 2008-2009**

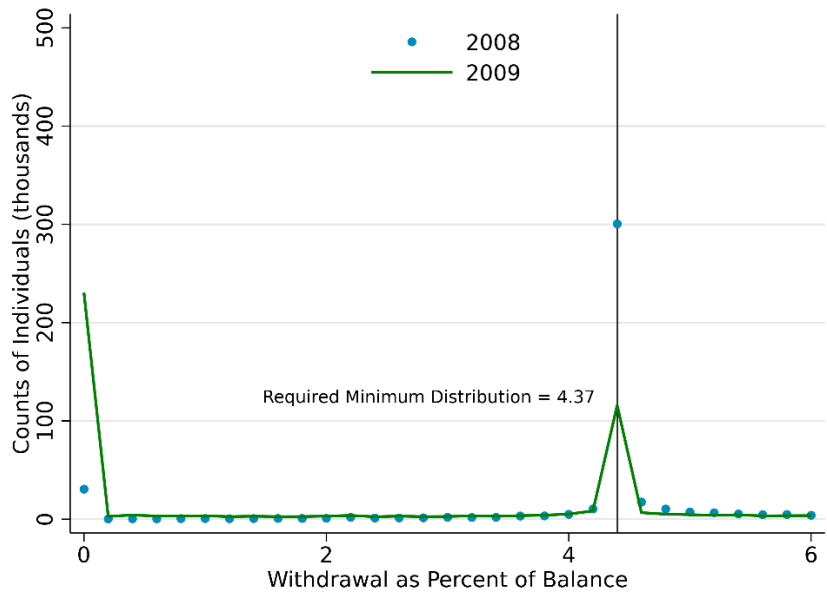

**Figure 4b. 2019-2020**

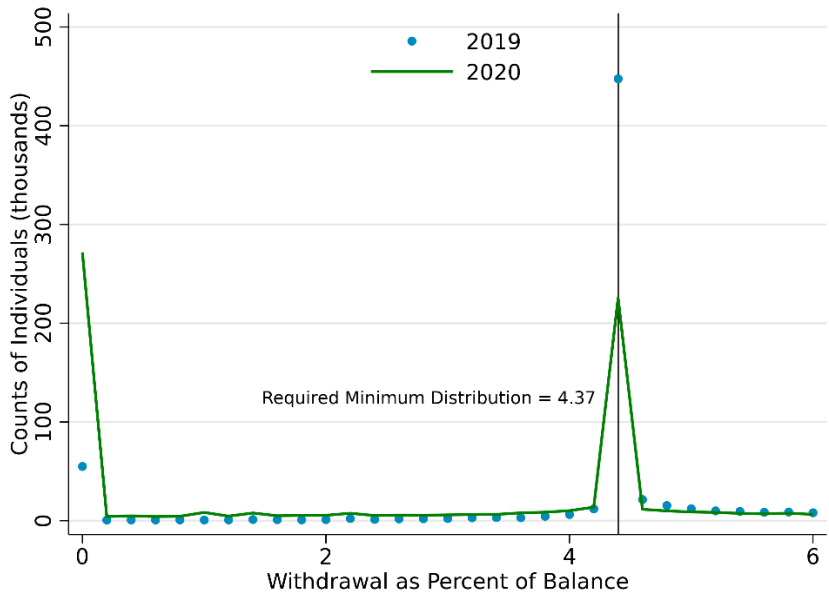

**Notes:** This figure plots a histogram of IRA withdrawals for 75-year-olds, scaled by prior year IRA balance, separately by year. In each panel, the blue dots represent the series in the years in which RMDs are in effect (2008 and 2019) and the green line represents the series in years in which RMDs have been suspended (2009 and 2020). The vertical line indicates the standard RMD amount for 75-year-olds. Withdrawals are derived from Form 1099-R and exclude rollovers and Roth conversions. The data underlying the figure are a 5 percent random sample of IRS records derived from tax returns and information returns.

**Figure 5: Total Withdrawals and Penalized Withdrawals to those under Age 59.5**

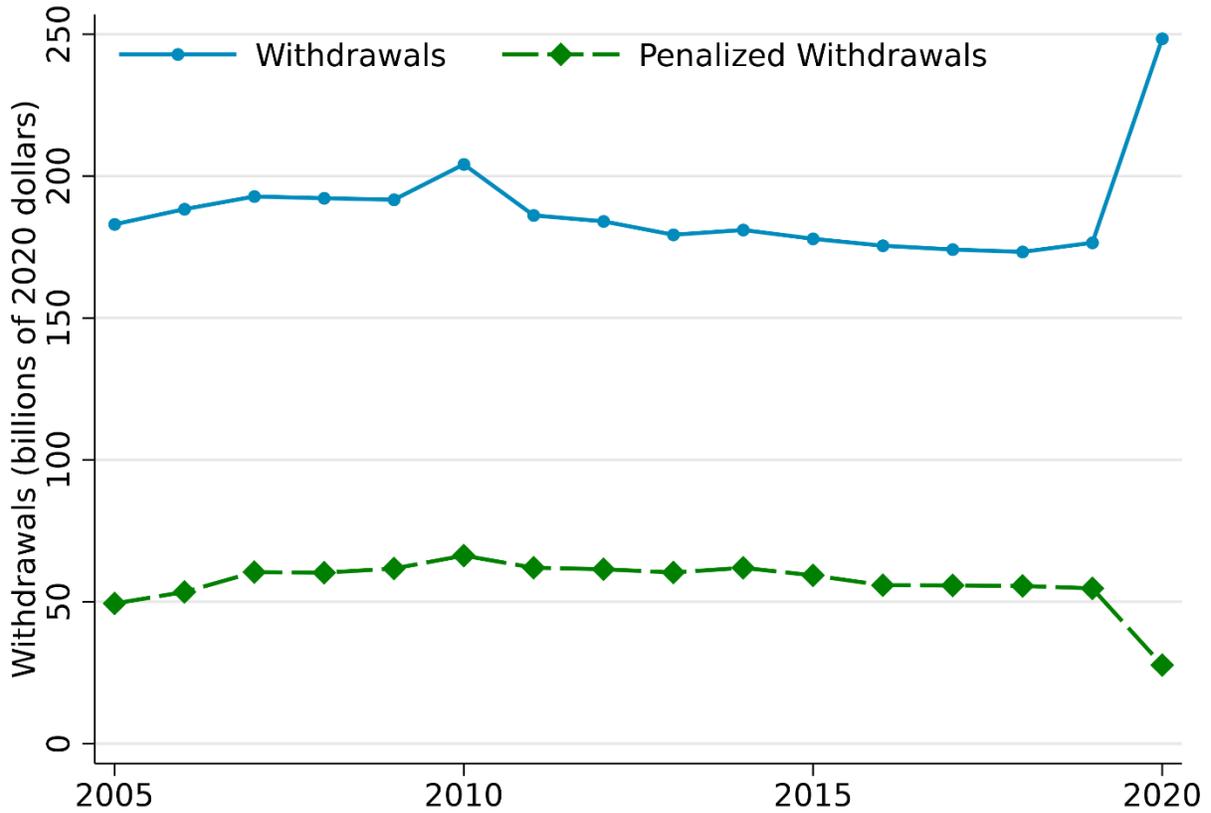

**Notes:** This figure plots withdrawals and (estimated) penalized withdrawals made to tax units with at least one member under age 59.5. Withdrawals are derived from Form 1099-R and exclude rollovers and Roth conversions. We estimate penalized withdrawals as equal to the penalty amount on Form 1040 divided by 0.1. All dollar amounts are adjusted to 2020 price levels. The data underlying the figure are a 5 percent random sample of IRS records derived from tax returns and information returns.

**Figure 6: Total Retirement Withdrawals Made near $100,000, 2019-2020**

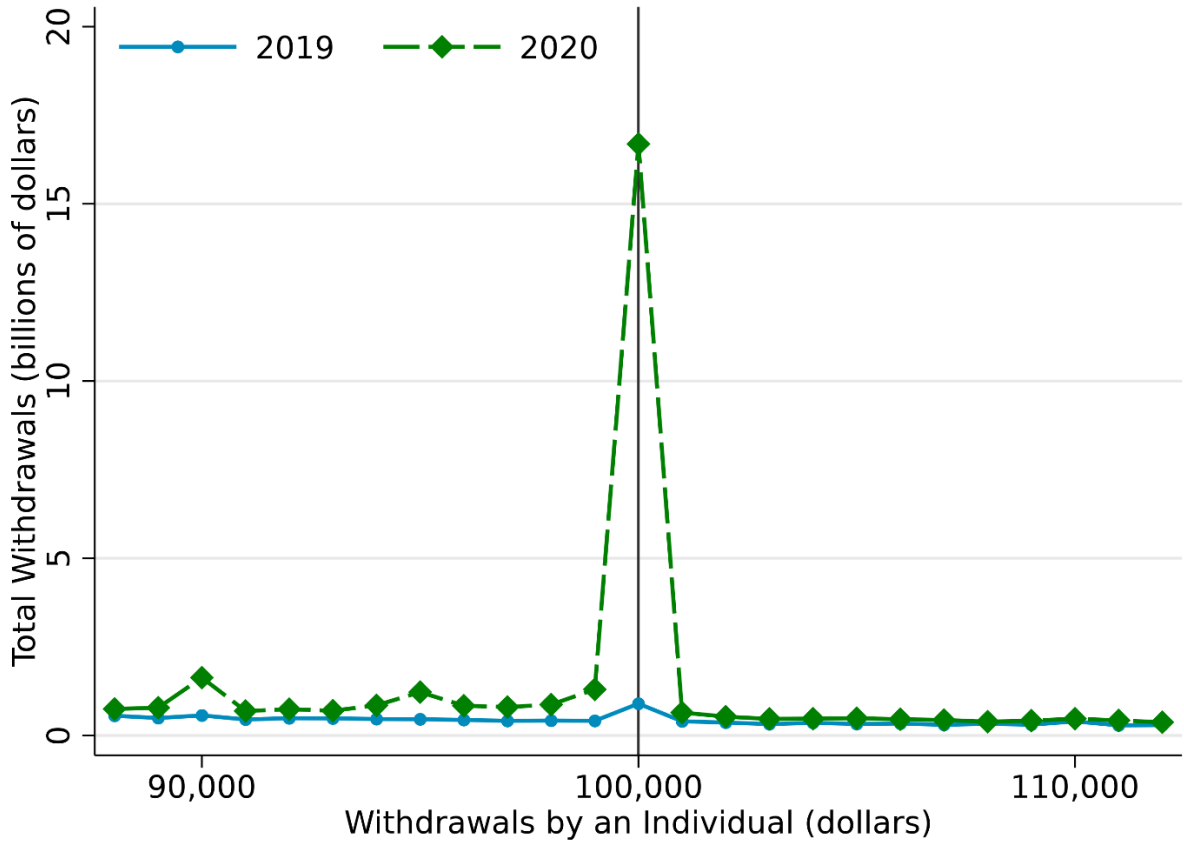

**Notes:** This figure plots a histogram of total withdrawals from IRAs and workplace plans made to individuals, broken out by the dollar amount of withdrawals made by an individual. The horizontal axis are thousand-dollar wide bins, and the vertical axis displays the total dollars of withdrawals made within each bin. Withdrawals are measured on Form 1099-R and exclude rollovers and Roth conversions. Only individuals between the ages of 20 and 58 are included. The data underlying the figure are a 5 percent random sample of IRS records derived from tax returns and information returns.

**Figure 7: Probability of Taking Withdrawals by Monthly Age near 59½**

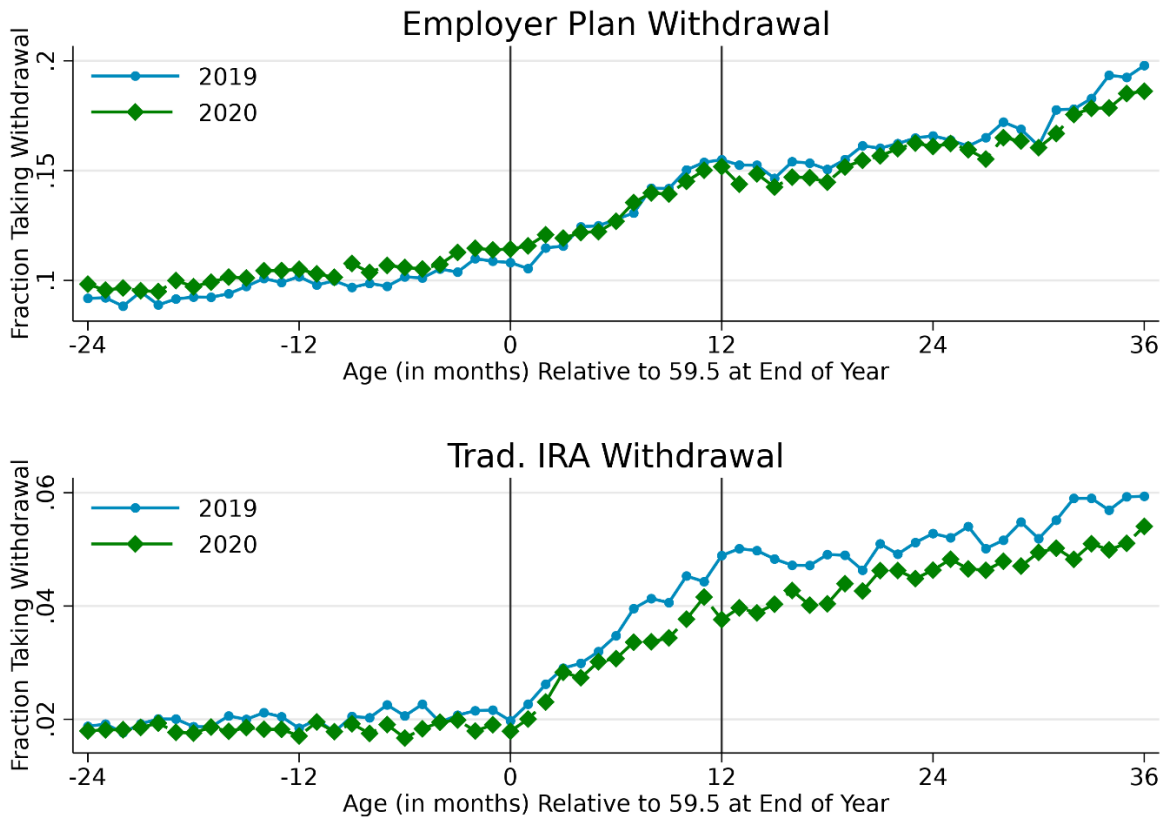

**Notes:** This figure plots the mean value of a dummy for taking any withdrawal from a workplace retirement plan (top panel) or IRA (bottom panel) as a function of monthly age, relative to 59½, separately by year. An individual with monthly age one attains 59½ in December of the given year, while an individual with monthly age zero attains 59½ in January of the subsequent year. Withdrawals are derived from Form 1099-R and exclude rollovers and Roth conversions. All dollar amounts are adjusted to 2020 price levels. The data underlying the figure are a 5 percent random sample of IRS records derived from tax returns and information returns.